\documentclass[showpacs,twocolumn,preprintnumbers,amsmath,amssymb]{revtex4}
\usepackage{amsmath,amsfonts,latexsym,amssymb,graphicx,graphics,epsfig,subfigure,color,makeidx}
\usepackage{xcolor,diagbox}
\usepackage{multirow}
\usepackage[colorlinks,linkcolor=blue,anchorcolor=blue,citecolor=green]{hyperref}
\newcommand {\nn}{\nonumber}
\begin{document}
\title{Generalized geometrical coupling for vector field localization {on thick brane in asymptotic Anti-de Sitter spacetime}}

\author{Tao-Tao Sui$^{a}$\footnote{suitt14@lzu.edu.cn}}
\author{Wen-Di Guo$^{a}$\footnote{guowd16@lzu.edu.cn}}
 \author{Qun-Ying Xie$^{b}$\footnote{xieqy@lzu.edu.cn}}
\author{Yu-Xiao Liu$^{a}$$^{c}$\footnote{liuyx@lzu.edu.cn, corresponding author}}

\affiliation{$^{a}$Institute of Theoretical Physics $\&$ Research Center of Gravitation, Lanzhou University, Lanzhou 730000, China\\
             $^{b}$School of Information Science and Engineering, Lanzhou University, Lanzhou 730000, China\\
            $^{c}$Key Laboratory for Magnetism and Magnetic of the Ministry of Education, Lanzhou University, Lanzhou 730000, China}

\begin{abstract}
It is known that a five-dimensional free vector field $A_{M}$ cannot be localized on   Randall-Sundrum (RS)-like thick branes,  namely, the thick branes embedded in asymptotic Anti-de Sitter (AdS)  spacetime. In order to localize a vector field on the RS-like thick brane, an extra coupling term should be introduced. In this paper, we generalize the geometrical coupling  mechanism  by adding two mass terms
($\alpha Rg^{MN}A_{M}A_{N}+\beta R^{MN}A_{M}A_{N}$) into the action. We decompose the  fundamental vector field $A_{M}$ into three parts: transverse vector part $\hat{A}_{\mu}$, scalar parts $\phi$ and $A_{5}$. Then, we find that the transverse vector  part $\hat{A}_{\mu}$  decouples from  the scalar parts. In order to eliminate the tachyonic modes of $\hat{A}_{\mu}$, the two coupling parameters $\alpha$ and
$\beta$ should satisfy a relation. {Combining the restricted condition, we can get a combination parameter as $\gamma=\frac{3}{2}\pm\sqrt{1+12\alpha}$.} Only if $\gamma>1/2$, the zero mode of $\hat{A}_{\mu}$ can be localized on the RS-like thick brane. We also investigate the resonant character of the vector part $\hat{A}_{\mu}$ for the general RS-like thick brane with the warp factor $A(z)=-\ln(1+k^2z^2)/2$ by choosing the relative probability method. The result shows that, only for $\gamma>3$, the massive resonant Kaluza-Klein modes can exist. The number of resonant Kaluza-Klein states increases with the combination parameter $\gamma$, and the lifetime of the first  resonant state can be long enough as the age of our universe.  This indicates that the vector resonances might be considered as one of the candidates of dark matter.  Combining the conditions of experimental observations, the constrain shows that the parameter $k$ has a lower limit with $k\gtrsim{10^{-17}}$ eV, the combination parameter $\gamma$ should be greater than $57$, and accordingly, the mass of the first resonant state should satisfy $m_{1}\gtrsim{10^{-15}}$ eV.
\end{abstract}


\maketitle

\section{Introduction}

Over the last decades, brane world theories have received a lot of attention for the success on solving the gauge hierarchy and cosmological constant problems \cite{Randall,Arkani-Hamed98}.  In the brane world scenario, our universe is {a 3-brane embedded} in a higher-dimensional bulk. The well known {Randall-Sundrum (RS)} models \cite{Randall} ({{including} RS-1 and RS-2 models}) involve one extra dimension with a non-trivial warp factor due to the underlying anti-de Sitter (AdS) geometry.

{In the RS thin brane models and their generalizations}, {the branes have no thickness and there are no} {dynamical mechanisms responsible for their formation}. {In order to investigate the dynamical generation of branes} {and their internal structure,} {domain wall (or thick brane) models were presented}, and for more details on thick brane models, see Refs. \cite{Dzhunushaliev:2009va,1707.08541}. {One of the features of thick brane is that} {usually it}  is generated by one or {more} background scalar fields coupled with gravity.

In thick brane models, various fundamental matter fields are living in the higher-dimensional bulk. Therefore, in order to construct a more realistic brane world, on which the four-dimensional gravity and matter fields in the standard model {should be localized, it} is very necessary and significant to provide an effective localization mechanism for the bulk gravity and matter fields.  The results of Refs.~
\cite{Gremm:1999pj,yizhong17,Barbosa-Cendejas,Farakos2007,GermanHerrera2012,
Herrera-A2012,Kakushadze:2000zp} show that four-dimensional gravity can be localized on the thick branes {generated by} background scalar field(s) in a five-dimensional {asymptotic AdS spacetime.} As shown in Refs.~\cite{ Linares,Liu7911,Koroteev08,Flachi09,Bajc2000,
Fuchune}, {a free massless scalar field} can also be localized on the thick {branes}. For {a Dirac fermion field}, without introducing the scalar-fermion coupling (also called the Yukawa coupling) \cite{ThickBrane2,guo1408,LiuXu2014,ThickBrane1,ThickBrane3,
Liu0803,Neupane} or fermion-gravity coupling \cite{LiLiu2017a}, {it has no} normalizable zero mode in five-dimensional RS-like brane models. Unfortunately, Ref.~\cite{Bajc:1999mh} {gave} the essence of  ``no-go theorem'' in the thin brane limit (RS-2 model with {an infinite} extra dimension) that the localization for {a vector field} seems to require a richer brane structure, {for examples}, the de Sitter brane \cite{Liu200808,Liu20090902,GuoHerrera2013,Herrera-Aguilar2014}, the brane world with finite extra dimension \cite{LiuFuGuoLi2012}, or a six-dimensional {string-like model} \cite{Oda2000}.

Then, a lot of works have been devoted to find a mechanism for vector field localization, and {many} literatures show a wide variety of ideas. Kehagias and Tamvakis proposed a dilaton coupling between the vector field and background scalar field \cite{Kehagias2001}. This mechanism has been widely applied {in different} thick brane models \cite{CruzTahimAlmeida2010,Alencar2010,
CruzLimaAlmeida2013,FuLiuGuo2011,CruzTahimAlmeida2009,ChristiansenCunhaTahim2010,
CruzMalufAlmeida2013}. After that, Chumbes, {Holf da Silva} and Hott proposed {a coupling function between the vector field and the background scalar field \cite{ChumbesHoffHott2012}.} Vaquera-Araujo and Corradini introduced a Yukawa-like coupling, namely, a Stueckelberg-like action, to realize the localization of {the vector field}  \cite{Vaquera-AraujoCorradini2014}.

{Recently, Zhao et al.} \cite{zhao1406} presented another localization mechanism of the vector field, {i.e., introduced} a mass term $\alpha Rg^{MN}A_{M}A_{N}$ with $R$ the five-dimensional scalar curvature. They found that only for {a special} coupling parameter $\alpha=-1/16$, the vector part $\hat{A}_{\mu}$ can be localized on the thick brane, and there are no tachyonic modes.  Then, Alencar et al. introduced other forms of the mass term: $\beta R^{MN}A_{M}A_{N}$ and $\beta G^{MN}A_{M}A_{N}$  with $R^{MN}$ and $G^{MN}$ the Ricci tensor and Einstein tensor \cite{Alencar1506}.
While, in all these mechanisms, in order to eliminate tachyonic vector modes, the massive parameters $\alpha$ or $\beta$ should be fixed, since there is no more degree of freedom for the coupling parameter. As a result, the effective potential {of the vetor Kaluza-Klein (KK) modes} is fixed and usually there are no resonant
vector KK modes quasi-localized on the brane.

Inspired by the above works, we generalize the mass term to the following one
\begin{eqnarray}
 {  -\frac{1}{2}\left(\alpha Rg^{MN}A_{M}A_{N}+\beta R^{MN}A_{M}A_{N}\right)}, \label{generalizedCoupling}
\end{eqnarray}
{since both terms are possible couplings. Then we} study the localization and quasi-localization of the vector field on the thick brane. For the quasi-localized massive KK modes, they {might be a candidate} for dark matter. {Note that, the consistency conditions for this kind of localization mechanism is just investigated by \cite{Freitas:2020mxr}.}

We decompose the vector field $A_{M}$ into three parts: {the transverse component $\hat{A}_{\mu}$ (transverse vector part), the longitudinal component $\partial_{\mu}\phi$ (scalar part), and the fifth component $A_{5}$ (scalar part). Here} the Latin indices ($M, N=0,1,2,3,5$) stand for the five-dimensional coordinate indices,  Greek indices ($\mu, \nu=0,1,2,3$) correspond to the brane coordinate indices. We find that the {transverse vector part} $\hat{A}_{\mu}$ decouples with the scalar parts $\phi$ and $A_{5}$. Besides, in order to eliminate the tachyonic modes of $\hat{A}_{\mu}$,  the two parameters in the coupling term \eqref{generalizedCoupling}, {$\alpha$ and $\beta$, should satisfy a relation:} $\beta=-1-8\alpha\pm\sqrt{1+12\alpha}$.  With this {constraint}, we can get {a combination parameter} $\gamma=\frac{3}{2}\pm\sqrt{1+12\alpha}$, and the localized condition for the {transverse vector part} $\hat{A}_\mu$ is $\gamma>1/2$. More importantly, we can find the resonant states under this restrict condition. We investigate the resonant character of $\hat{A}_\mu$ with the general RS-like thick brane warp factor $A(z)=-\ln(1+k^2z^2)/2$. These resonant states can be considered as {a candidate} for dark matter.

The remaining parts are organized as follows. In section \ref{sec-2}, we introduce the generalized model of {the vector field}. Then, we calculate the localization of {the transverse part of a five-dimensional vector} on the thick brane in section \ref{localization}. After that, we study the resonant character of the {transverse vector part} in section \ref{resonance}. Finally, we conclude our results in the last section.

\section{{The generalized geometrical coupling mechanism of vector field}}\label{sec-2}

{The vector field can be localized on the thick brane by considering the geometrical coupling {term, e.g., the coupling between the vector field and the Ricci scalar (or Ricci tensor), which} can be viewed as a mass term \cite{zhao1406,Alencar1409,Alencar1506}. In this paper, we consider the generalized geometrical coupling (\ref{generalizedCoupling}). Then the full five-dimensional action for the vector field $A_M$ {is given by}
\begin{eqnarray}
  S&=&-\frac{1}{4}\int d^5x\sqrt{-g}\Big( {F^{MN}F_{MN}} \nn\\
 &&+2(\alpha R g^{MN}+\beta R^{MN})A_M A_N\Big ),\label{vectorAction}
 \end{eqnarray}
where $g_{MN}$ is the metric of the five-dimensional bulk spacetime and $F_{MN}=\partial_{M}A_{N}
-\partial_{N}A_{M}$ is the field {strength. Here,} we decompose the vector $A_{M}$ in the following way:
\begin{equation}
A_{M}=(\hat{A}_{\mu}+\partial_{\mu}\phi,\;A_{5}), \label{eqAM}
\end{equation}
where $\hat{A}_{\mu}$ is the transverse component with the transverse condition $\partial_{\mu}\hat{A}^{\mu}=0$, and $\partial_{\mu}\phi$ is the longitudinal component.

We adopt the following metric ansatz to describe the five-dimensional spacetime:
\begin{equation}\label{metric}
ds^2=e^{2A(y)}\eta_{\mu\nu}dx^{\mu}dx^{\nu}+dy^2,
\end{equation}
where the warp factor $A(y)$ is a function of the extra dimensional coordinate $y$.
So,  {the Ricci scalar and the nonvanishing components of the Ricci tensor} can be expressed as
\begin{eqnarray}
R~&=&-4(5A'^2+2A''),\\
R^{\mu\nu} &=&-(4{A'}^2+A'')g^{\mu\nu},\\
R^{55} &=& -4({A'}^2+A''),
\end{eqnarray}
where the prime denotes derivative with respect to $y$. {The components of the mass
terms} {in action (\ref{vectorAction})} can the written as
 \begin{eqnarray}
\alpha R g^{\mu\nu}+\beta R^{\mu\nu} &=& \mathcal{W}g^{\mu\nu},\\
\alpha R g^{5 5}+\beta R^{5 5} &=& \mathcal{G}g^{5 5},
 \end{eqnarray}
{where}
 \begin{eqnarray}
\mathcal{W} &=& -4(5\alpha+\beta){A'}^2-(8\alpha+\beta)A'',\\
 \mathcal{G} &=& -4(5\alpha+\beta){A'}^2-4(2\alpha+\beta)A''.
 \end{eqnarray}

{By substituting} the decomposition \eqref{eqAM} into the action \eqref{vectorAction}, we can split the action into {two parts}
 \begin{equation}
  S=S_{V}(\hat{A}_{\mu})+S_{S}(\phi,\,A_5),
 \end{equation}
where \begin{eqnarray}
S_{V}&=&-\frac{1}{4}\int d^5x\Big(\hat{F}_{\lambda \mu}\hat{F}_{\nu \rho}{\eta}^{\lambda\nu} {\eta}^{\mu\rho}+2{\partial}_{5}{\hat{A}_{\mu}}{\partial}^{5}{\hat{A}_{\nu}}\eta^{\mu\nu}e^{2A}\nn\\
&+&2\mathcal{W}\hat{A}_{\mu}\hat{A}_{\nu}\eta^{\mu\nu}e^{2A}\Big), \label{actionT} \\
S_{S}&=&-\frac{1}{2}\int d^5x e^{2A}\Big(\eta^{\mu\nu}g^{55}({\partial}_{5}{\partial}_{\mu}{\phi})({\partial}_{5}{\partial}_{\nu}{\phi})\nn\\
&+&\mathcal{W}\eta^{\mu\nu} {\partial}_{\mu}{\phi}{\partial}_{\nu}{\phi}
+\eta^{\mu\nu}g^{55}{\partial}_{\mu}{A}_{5}{\partial}_{\nu}{A}_{5}\nn\\
&+&\mathcal{G}e^{2A}g^{55}{A}_{5}{A}_{5}-2\eta^{\mu\nu}g^{55}{\partial}_{\mu}{A}_{5}
(\partial_{5}{\partial}_{\nu}{\phi})\Big), \label{actionscalar}
\end{eqnarray}
{where $\hat{F}_{\mu\nu} = \partial_{\mu} \hat{A}_{\nu}-\partial_{\nu} \hat{A}_{\mu}$. The above result shows that the transverse vector part $\hat{A}_\mu$ decouples from the scalar parts. So, we only consider separately the localization condition and resonant character of the transverse vector part $\hat{A}_\mu$.

\section{Localization of the transverse vector part of the vector field on thick brane}
\label{localization}

In this section, we consider the localization of the vector part $\hat{A}_\mu$ independently. We make the following KK decomposition:
\begin{equation}
\hat{A}_{\mu}(x,y)=\sum_{n}a^{(n)}_{\mu}(x^\nu)\tilde{\rho}_{n}(y),\label{KKdeco}
\end{equation}
where $a^{(n)}_{\mu}(x^\nu)$ is the four-dimensional vector KK mode and $\tilde{\rho}_{n}(y)$ is the corresponding extra dimensional profile (also called as the KK wave function in Ref. \cite{Ponton2012}), and the index ``$n$'' represents the  $n$-th KK mode. By using the KK decomposition \eqref{KKdeco} and  the orthonormality condition
\begin{equation}\label{ortconditions}
\int^{\infty}_{-\infty}\tilde{\rho}_{n}(y)\tilde{\rho}_{m}(y)=\delta_{mn},
\end{equation}
we can get an effective action including the four-dimensional massless vector field (the zero mode $a^{(0)}_{\mu}$) and a set of massive vector fields $a^{(n)}_{\mu}$ with $n>0$:
\begin{equation}\label{actionT4}
S_{V}=-\frac{1}{4}\sum_{n}\int d^4x\Big(
f^{(n)}_{\mu\lambda}f_{(n)}^{\mu\lambda}+2 m_n^2 a^{(n)}_{\mu}a^{(n)}_{\nu}\eta^{\mu\nu}\Big),
\end{equation}
where $f^{(n)}_{\mu\nu}=\partial_{\mu}a^{(n)}_{\nu}-\partial_{\nu}a^{(n)}_{\mu}$ is the four-dimensional vector field strength tensor. In addition, the extra dimensional part $\tilde{\rho}_{n}(y)$ should satisfy the following equation:
\begin{equation}
-\partial_y\left(e^{2A(y)}\partial_{y}\tilde{\rho}_n\right)+\tilde{\rho}_{n}e^{2A(y)}\mathcal{W}
=m_n^2\tilde{\rho}_n. \label{eq}
\end{equation}

{In order to solve the above equation} \eqref{eq}, {we make} a coordinate transformation $dz=e^{-A(y)}dy$, {for which} the metric can be expressed as
\begin{equation}\label{conformalmetric}
ds^2=e^{2A(z)}(\eta_{\mu\nu}dx^{\mu}dx^{\nu}+dz^2),
\end{equation}
{and Eq. \eqref{eq} is rewritten} as
\begin{equation}
-\partial_z\Big(e^{A(z)}\partial_z\tilde{\rho}_{n}\Big)+\tilde{\rho}_{n}e^{3A(z)}\mathcal{W}
=e^{A(z)}m^2_{n}\tilde{\rho}_{n},\label{eq1}
\end{equation}
with $\mathcal{W}=e^{-2A}(-12\alpha-3\beta)(\partial_{z}A)^2+e^{-2A}(-8\alpha-\beta)\partial^2_{z}A$. {After the field transformation $\tilde{\rho}_{n}=e^{-\frac{1}{2}A(z)}\rho_{z}(z)$}, Eq. \eqref{eq1} can be rewritten as a Schr\"{o}dinger-like equation:
\begin{equation}
\Big(-\partial^{2}_{z}+V_{v}(z)\Big)\rho_{n}=m^2_{n}\rho_{n},\label{equa1}
\end{equation}
where the explicit expression of the effective potential $V_{v}(z)$ is
\begin{equation}\label{effpov}
V_{v}(z)=\left(\frac{1}{4}-12\alpha-3\beta\right)(\partial_{z}A)^2
         +\left(\frac{1}{2}-8\alpha-\beta\right)\partial^2_{z}A.
\end{equation}

In order to exclude the tachyonic vector modes, the {eigenvalues} of Schr\"{o}dinger-like equation \eqref{equa1} should be non-negative, i.e., $m_{n}^2\geq0$. So, Eq. \eqref{equa1}  should be written in the form of $Q^{+}Q\rho_{n}= m_{n}^{2}\rho_{n}$ with $Q=-\partial_{z}+\gamma
\partial_{z}A$. That is to say, the effective potential should be in the form of
\begin{eqnarray}
V_{v}(z)=\gamma^2(\partial_{z}A)^2+\gamma\partial^2_{z}A.
 \label{effpov2}
\end{eqnarray}
{To this end}, the two parameters $\alpha$ and $\beta$ should satisfy the following {relation:}
\begin{equation}
\beta=-1-8\alpha\pm\sqrt{1+12\alpha}, \label{relation}
\end{equation}
{and so parameter $\gamma$ in \eqref{effpov2} is given by}
\begin{eqnarray}
\gamma=\frac{3}{2}\pm\sqrt{1+12\alpha}.\label{gamma}
\end{eqnarray}
{With the relation \eqref{relation} and the expression \eqref{gamma},} the Schr\"{o}dinger-like equation \eqref{equa1} can be {further} rewritten as
\begin{equation}\label{sequ2}
\Big(-\partial^{2}_{z}+\gamma^2(\partial_{z}A)^2+\gamma\partial^2_{z}A\Big)\rho_{n}=m^2_{n}\rho_{n}.
\end{equation}

{Now, we investigate the localization of the zero mode of $\hat{A}_{\mu}$, for which $m_{0}=0$ and the solution is given by}
\begin{equation}
\rho_{0}(z)=c_{0}e^{\gamma A(z)},
\end{equation}
{where $c_{0}$ is the normalization constant.}
According to the orthonormality condition \eqref{ortconditions}, the integration of ${\rho}^2_{0}$ should be finite, namely,
\begin{eqnarray}
\int_{-\infty}^{+\infty}{\rho}^2_{0}dz&=&{c_0^2}\int_{-\infty}^{+\infty} e^{2\gamma A(z)}dz\nn\\
&=&{c_0^2}\int_{-\infty}^{+\infty} e^{(2\gamma-1)A(y)}dy =1  \label{int31}.
\end{eqnarray}
For the RS-like braneworld scenarios, the warp factor has the following asymptotic behavior as
\begin{equation}
A(y)|_{y\rightarrow\pm\infty}\rightarrow-k|y|,
\end{equation}
where $k$ is the scale parameter  of the brane with mass dimension. Plugging it into Eq. \eqref{int31}, we obtain that
\begin{equation}
{e^{(2\gamma-1)A(y)}|_{y\rightarrow\pm\infty} \rightarrow  e^{-(2\gamma-1)k|y|}.} \label{integration}
\end{equation}
{In order to ensure the integration \eqref{int31} is convergent, the parameter should satisfy $\gamma>1/2$, i.e., $2\pm\sqrt{1+12
\alpha}>0$.} So, the range of the parameter $\alpha$ for different concrete expressions of $\beta$ are:
\begin{eqnarray}\label{locondition}
 \alpha>-1/12, \beta&=&-1-8\alpha-\sqrt{1+12\alpha},\label{condition1}\\
  0>\alpha>-1/12,  \beta&=&-1-8\alpha+\sqrt{1+12\alpha}.\label{condition2}
\end{eqnarray}

\section{The resonant character of $\hat{A}_{\mu}$} \label{resonance}
{In this section, we would like to investigate the massive KK states of the transverse vector part for the vector field. We will mainly look for resonant KK states of the vector field, which are quasi-localized on the brane but will propagate into extra dimension eventually. The resonance spectrum of these KK states is one of the typical characteristics of RS-like brane world models. They can interact with four-dimensional particles, which may led to the non-conservation of energy and momentum since the KK resonances can escape out of the brane. So, it is} possible to probe extra dimensions by detecting resonant states \cite{Aaltonen}. Besides, some physicists regard those massive KK particles as a candidate of dark matter (see Refs.~\cite{KK1,KK2,KK3} the details). The appearance of these resonances is related to the structure of the brane. Thus, it is important and interesting to study the resonant KK modes on the thick brane with different structures. References \cite{Almeida1,fuqumi,Fermion1,Cruz1,Landim,
duyuzhi} have considered resonances of graviton and fermion. {Besides,  Arakawa} et al. considered a massive vector field as a candidate of dark matter to explain the strong CP problem \cite{tait19}. So, we will study the {resonances} of {the five-dimensional vector field}.

In order to study the resonant states, Almeida et al. proposed the large peaks of the wave function as the resonances method to study the fermion resonances \cite{Almeida1}. Then, Landim et al. {researched the resonant states with transfer matrix method \cite{Landim,duyuzhi}.} Here, we will choose the relative probability method proposed by Liu et al. \cite{Fermion1} to calculate the resonant KK modes of the vector part $\hat{A}_\mu$ {since the method is effective for both odd and even KK states}. The relative probability is defined as \cite{Fermion1}
\begin{equation}\label{relative probability}
P=\frac{\int^{z_b}_{-z_b}|\rho_{n}(z)|^2dz}{\int^{z_{max}}_{-z_{max}}|\rho_{n}(z)|^2dz},
\end{equation}
where $2z_b$ is approximately the width of the thick brane and $z_{max}=10z_b$. Since the potentials considered in this paper are symmetric, the wave functions are either even or odd. Hence, we can use the following boundary conditions to solve the differential equation \eqref{sequ2} numerically:
\begin{eqnarray}
\rho_{n}(0)&=& 0,~~\rho'_{n}(0)=1,~~~\text{for odd KK modes}, \nn \\
\rho_{n}(0)&=& 1,~~\rho'_{n}(0)=0,~~~\text{for even KK modes}.
\end{eqnarray}

We solve the {Schr\"{o}dinger}-like equation (25) with the general RS-like warp factor $A(z)=-\ln(1+{k^2z^2})/2$. According to the supersymmetric quantum mechanics, the supersymmetric {partner} potentials will share the same spectrum of massive excited states. So, we can judge whether there are resonances by analyzing the shape of the supersymmetric partaner potential (we call it the dual potential). For our case, the dual potential corresponding to \eqref{effpov2} is $V_{v}^{(\text{dual})}(z)=\gamma^2(\partial_{z}A)^2-\gamma\partial^2_{z}A$. If there is no well or quasi-well in the dual potential, then there is no resonances. Thus, only for $\gamma>3$, there might exist resonances. We solve the {KK states numerically. The result} shows that both the parameters $k$ and $\gamma$ will affect the properties of the resonant states.

Figure \ref{figure1} shows the influence of the combination parameter $\gamma$ on the effective potential $V_{v}(z)$ and the resonant KK modes of the vector field $\hat{A}_{\mu}$.  Figure \ref{potentialla} shows that the height of the potential  {barrier} increases with the combination parameter $\gamma$,  {which indicates} there are more resonant KK modes for larger $\gamma$, and this can be  {confirmed} from Figs. \ref{resonantv1}, \ref{resonantv2}, \ref{resonantv3}. Combining Figs. \ref{resonantv1}, \ref{resonantv2}, \ref{resonantv3}, we can see that the mass of the first resonant KK modes, the number of resonant states, and the mass gap of the resonant KK modes increase with the parameter $\gamma$.

The effect of the scale parameter $k$ is shown in Fig. \ref{figure2}. From Fig. \ref{potentialk}, we can see that the scale parameter $k$ can influence not only the width of the potential well, but also its  {height}.  {With the increasing of the scale parameter $k$}, the potential well becomes narrower and higher.  {From} Figs. \ref{resonantk1}, \ref{resonantk2}, \ref{resonantk3}, we can see that the mass of the first resonant KK mode and the mass gap of the resonant KK modes increase with the parameter $k$.  {However, the number of the resonances does not change with $k$ for a fixed $\gamma$.}
\begin{figure}[htb]
\subfigure[$V_{v}$]{\label{potentialla}
\includegraphics[width=0.22\textwidth]{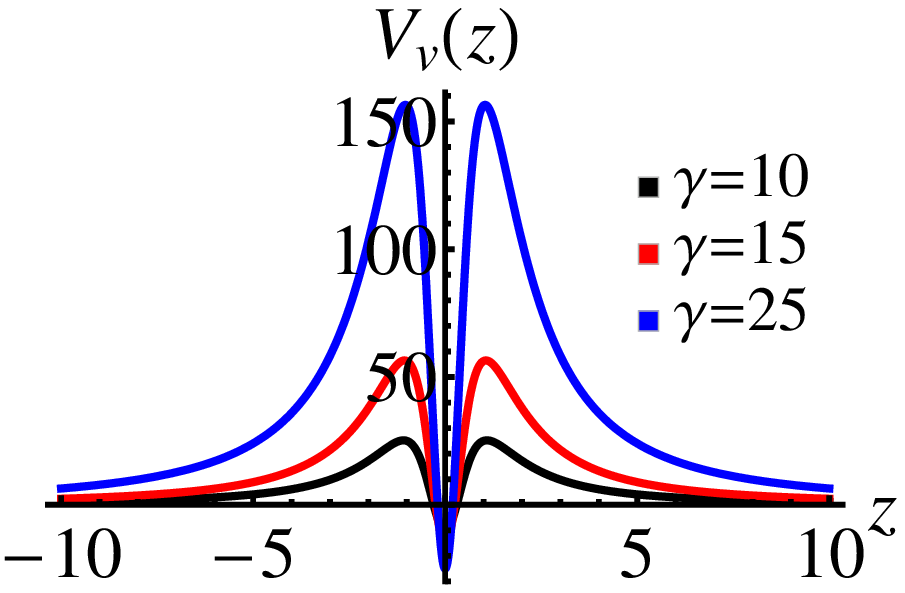}}
\subfigure[$\gamma=10$]{\label{resonantv1}
\includegraphics[width=0.22\textwidth]{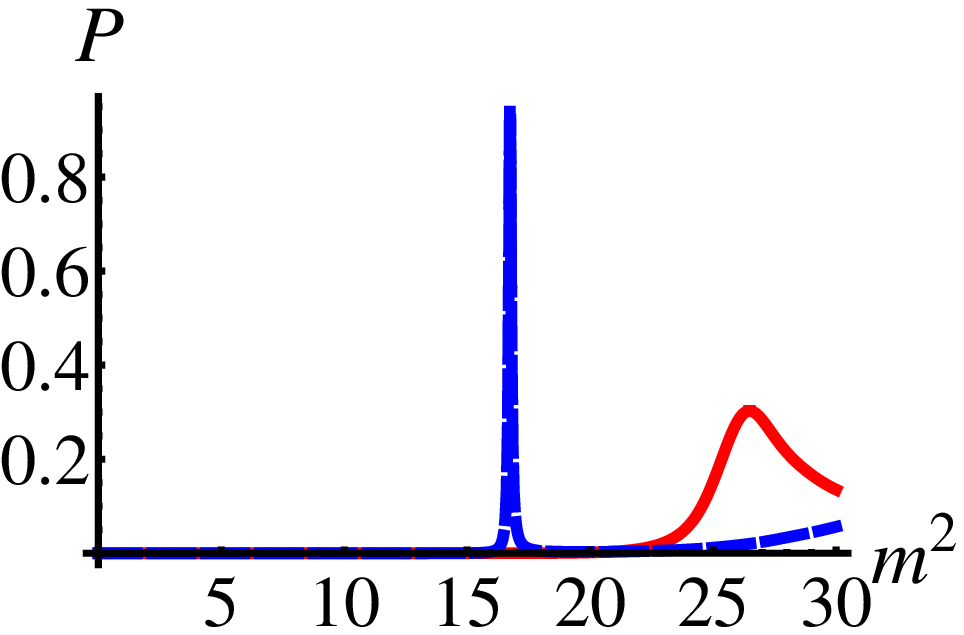}}
\subfigure[$\gamma=15$]{\label{resonantv2}
\includegraphics[width=0.22\textwidth]{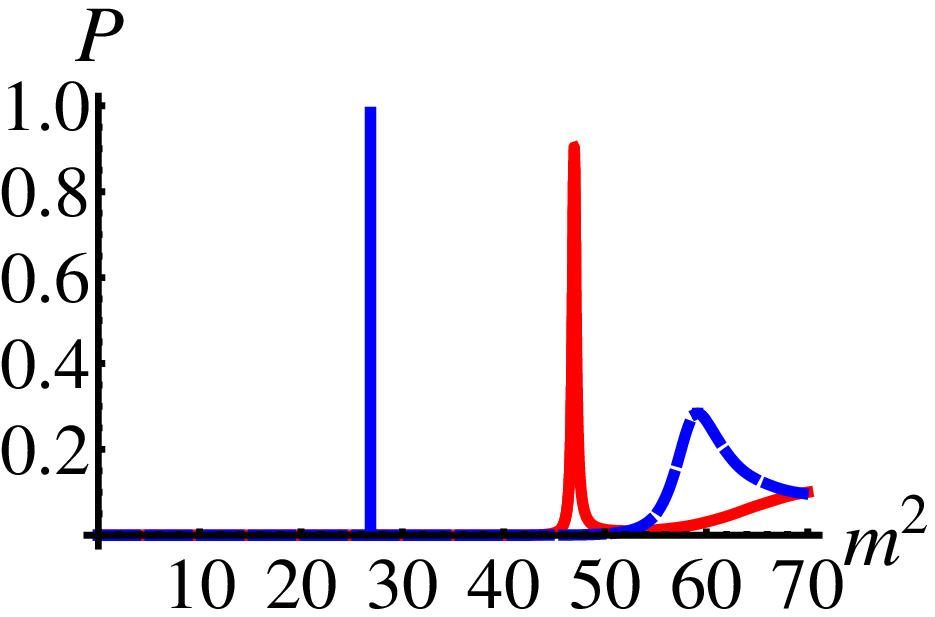}}
\subfigure[$\gamma=25$]{\label{resonantv3}
\includegraphics[width=0.22\textwidth]{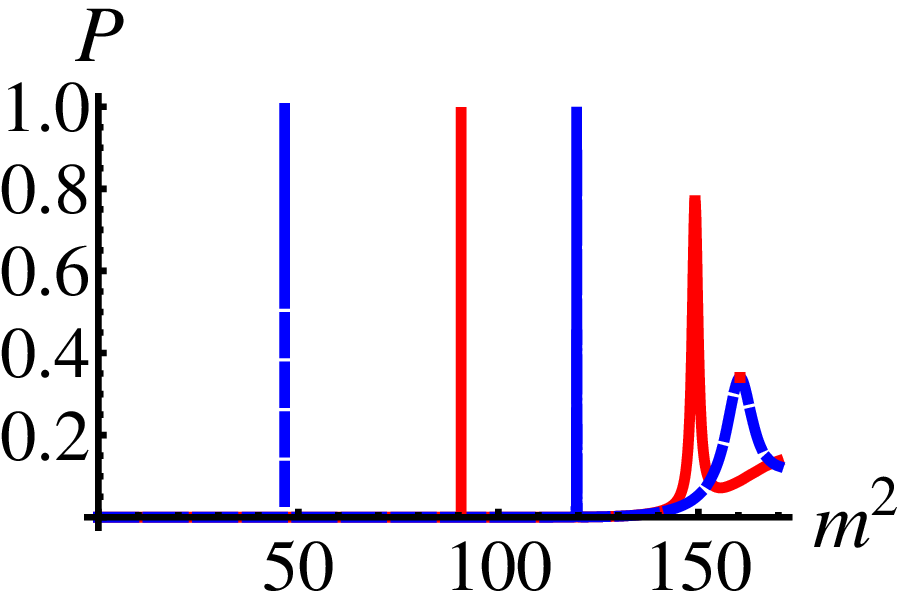}}
\caption{The influence of the combination parameter $\gamma$ on the effective potential $V_{v}$ and the probabilities $P$ (as a function of $m^2$)  {for the} odd-parity (blue dashed lines) and even-parity (red lines) massive KK modes. The scale parameter is set as $k=1$.}
\label{figure1}
\end{figure}
\begin{figure}[htb]
\subfigure[$V_{v}$]{\label{potentialk}
\includegraphics[width=0.22\textwidth]{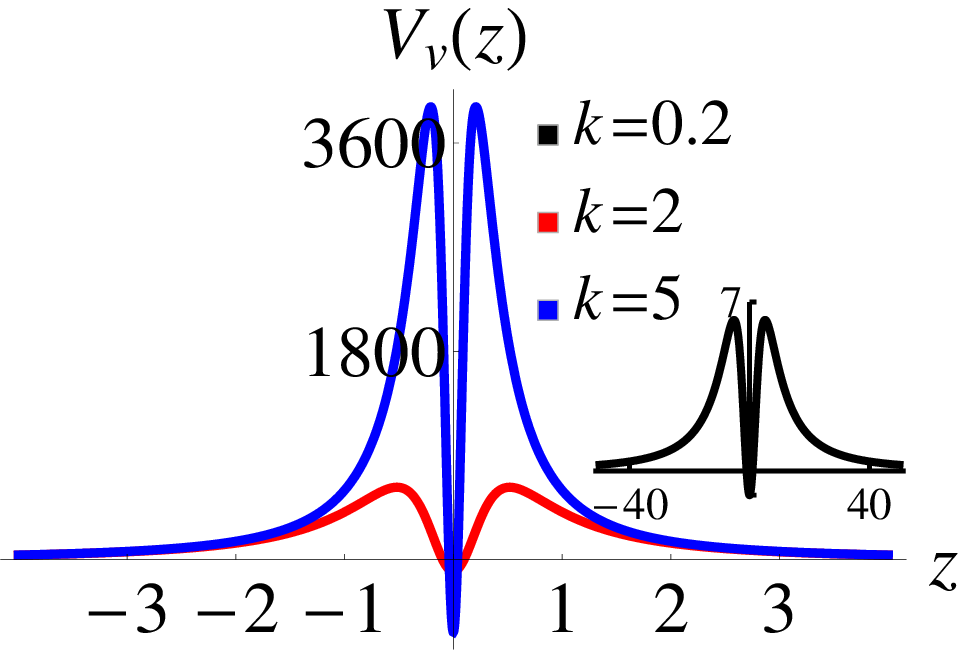}}
\subfigure[$k=0.2$]{\label{resonantk1}
\includegraphics[width=0.22\textwidth]{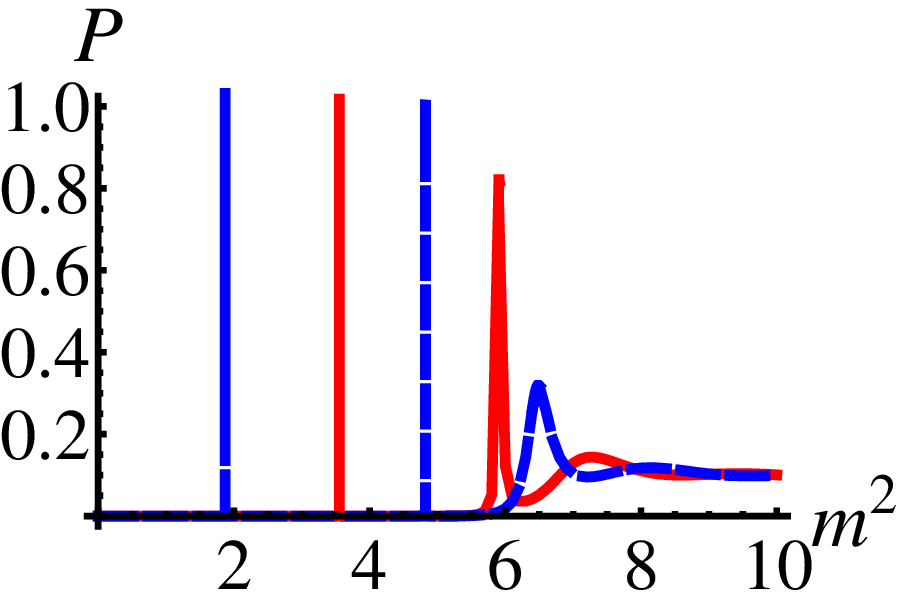}}
\subfigure[$k=2$]{\label{resonantk2}
\includegraphics[width=0.22\textwidth]{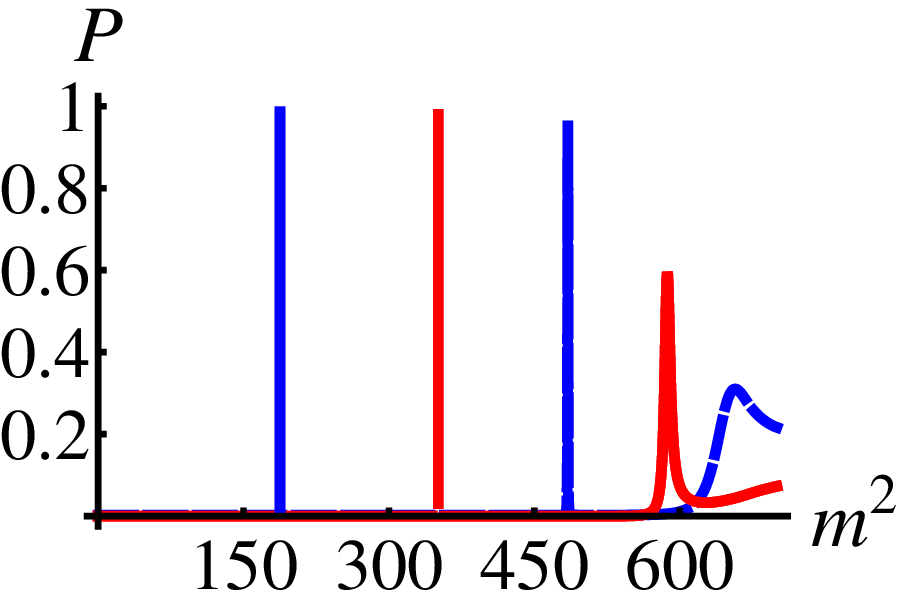}}
\subfigure[$k=5$]{\label{resonantk3}
\includegraphics[width=0.22\textwidth]{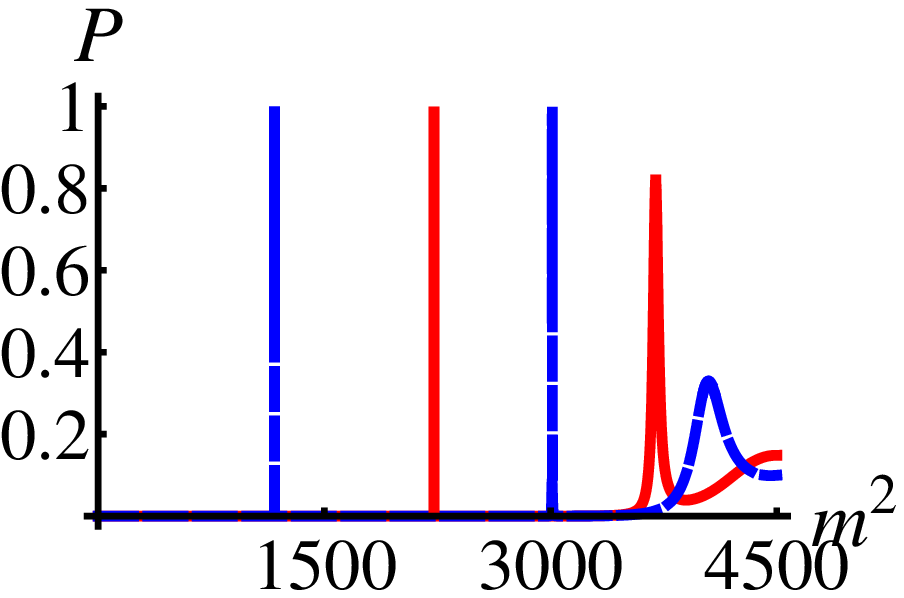}}
\caption{The influence of the scale parameter $k$ on the effective potential $V_{v}$ and the probabilities $P$ (as a function of $m^2$)  {for both} the odd-parity (blue dashed lines) and even-parity (red lines) massive KK modes. The combination parameter is set as $\gamma=25$.}
\label{figure2}
\end{figure}

Tables \ref{Table2} and \ref{Table1} are the specific values of the mass $m_n$, the relative probability $P$, the width $\Gamma$, and the lifetime $\tau$ for different parameters. Here, we define the width $\Gamma=\Delta m_n$  at half maximum of the peak and $\tau=1/\Gamma$. Table \ref{Table2} shows that  {with the increasing of} the parameter $\gamma$, the relative probability $P$ of the corresponding $n$-th resonant state becomes larger, and the lifetime of the resonant state becomes longer. From Tab.~\ref{Table1}, we can see that when the parameter $\gamma$ is fixed, all of the mass $m_{n}$, the width $\Gamma$, and the lifetime $\tau$ for the corresponding $n$-th resonant state are influenced by the parameter $k$. However, the values of $m_n/k$ are basically  {the same} for different values of $k$,  {so do the relative probability} $P$, the relative width $\Gamma/k$, and the relative lifetime $\tau*k$. So we can make a coordinate transformation as $\bar{z}=kz$ to offset the effect of $k$. Combining these two tables, we can see that the lifetime $\tau$ increases with $\gamma$, while decreases with $k$, which means that if the parameter $\gamma$ is large enough or the parameter $k$ is small enough, the lifetime of the resonant states can be long enough as the age of our universe. So, in this case, we can consider the resonant states as one of the candidates of dark matter.

Then, we calculate the lifetime of the first resonant state in order to check whether it can be a candidate of dark matter or not.  For convenience, we make a coordinate transformation $\bar{z}=kz$, and define the scaled mass $\bar{m}_1=m_1/k$ and the scaled lifetime $\bar{\tau}=1/\bar{\Gamma}$ for the first resonant state in the Natural System of Units, where $\bar{\Gamma}=\Delta \bar{m}_1$ is the half maximum of the peak for the first resonant state. Note that, both $\bar{m}_{1}$ and $\bar{\tau}$ are dimensionless.

Figure \ref{canshutu} shows that both the scaled mass $\bar{m}_1$ and the scaled lifetime $\text{log}(\bar{\tau})$ {linearly depend on the} parameter $\gamma$, and the fit functions can be expressed as
\begin{eqnarray}
\bar{m}_1&=&-3.2+2.0\gamma,\label{fitm}\\
\text{log}(\bar{\tau})&=&4.7+0.2\gamma.\label{fitt}
\end{eqnarray}

 \begin{figure}[htb]
\subfigure[$\bar{m}_1$]{\label{gammaM}
\includegraphics[width=0.22\textwidth]{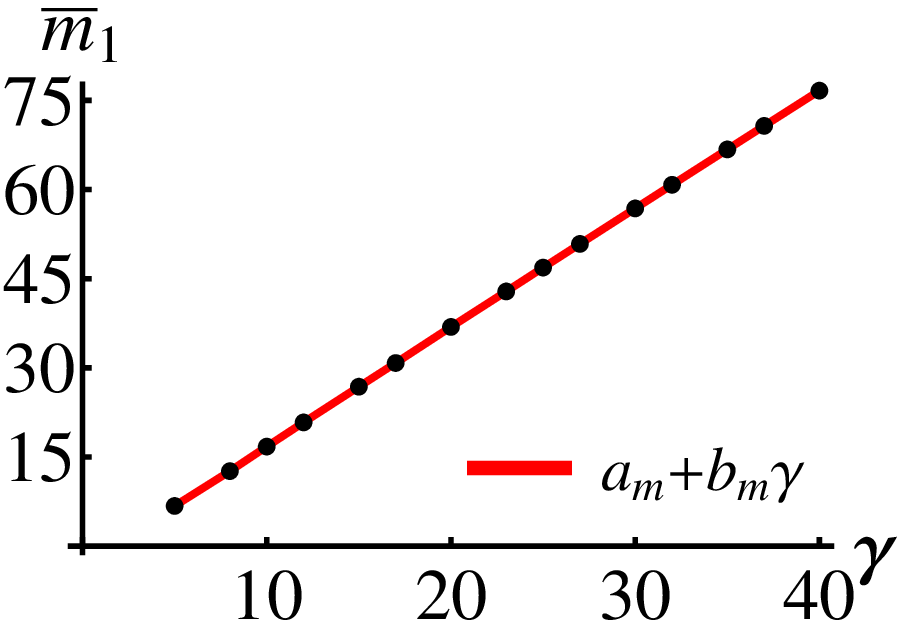}}
\subfigure[$\text{log}(\bar{\tau})$]{\label{gammaT}
\includegraphics[width=0.22\textwidth]{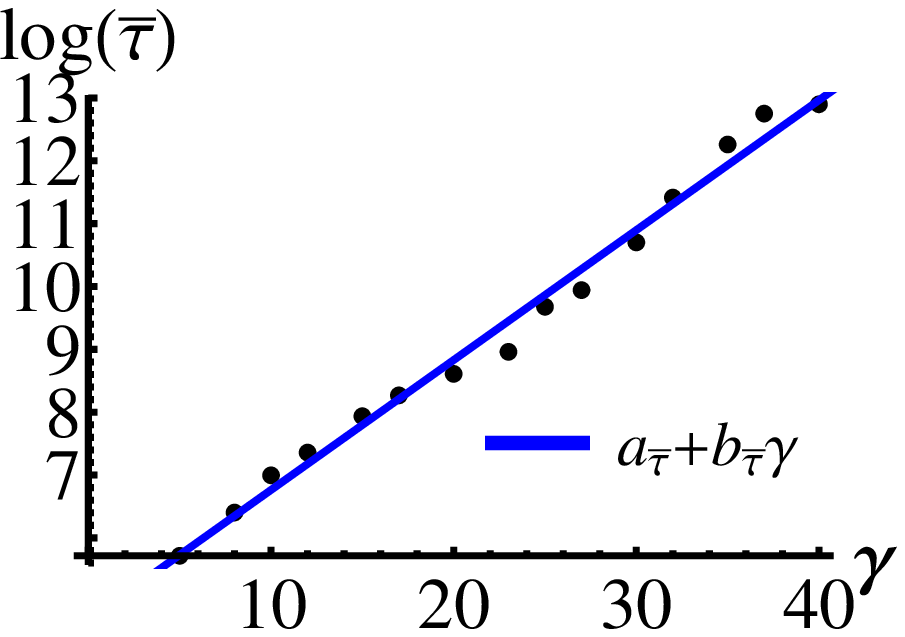}}
\caption{The influence of the combination parameter $\gamma$ on the scaled mass $\bar{m}_1$ and the scaled lifetime $\text{log}(\bar{\tau})$ of the first resonant state. The black dots are  {numerical results}, the red solid line is the fit function for $\bar{m}_{1}$ with $a_m=-3.2$ and $b_m=2.0$, and the blue solid line is the fit function for log($\bar{\tau}$) with $a_{\bar{\tau}}=4.7$, $b_{\bar{\tau}}=0.2$.}
\label{canshutu}
\end{figure}

{ It is known that} the age of our universe { is of about 13.8} billion years, i.e., { $4.35\times10^{17} \text{s}$}. So, if we consider { the} first resonant state as one of candidates of dark matter, { its lifetime} should be { larger than the age of universe}, i.e., { $\tau \gtrsim 4.35 \times 10^{17} \text{s}$, or in the Natural System of Units, }
 \begin{equation}\label{conditiontau}
\tau=1/(k\bar{\Gamma})=\bar{\tau}/k \gtrsim 6.6\times10^{32}\text{eV}^{-1}.
 \end{equation}
Thus, the restriction of the scale parameter $k$ can be expressed as
 \begin{equation}
 k \lesssim 1.5 \times{10^{-33}} \bar{\tau}~\text{eV}\simeq 7.5 \times 10^{-29+0.2\gamma}~\text{eV}.\label{conditionkgamma}
 \end{equation}
In addition, in { the brane world theory considered in this paper}, the relation between the four-dimensional effective Planck scale $M_{Pl}$ and the five-dimensional fundamental scale $M_{*}$ is given by \cite{Kakushadze:2000zp}:
 \begin{equation}
M_{Pl}^2=M_{*}^3\int_{-\infty}^{\infty} dz e^{3A(z)}
       =2M_{*}^3/k. \label{conditionM}
\end{equation}
Theoretically, if the energy scale {  reaches} the five-dimensional fundamental scale $M_{*}$, the quantum effect of gravity cannot be ignored. Experimentally, in the recent experiment of the Large Hadron Collider (LHC), { the collision energy is 13 TeV} and the result shows that the quantum effect of gravity can be ignored, which means that the five-dimensional fundamental scale $M_{*}>13$ TeV. 
{ Thus, the constrain on the parameter $k$ is}
 \begin{equation}\label{conditionM}
 k > 4.4\times{10^{-17}}~\text{eV}.
 \end{equation}

By combining the two conditions \eqref{conditionkgamma}, \eqref{conditionM} and the fit function \eqref{fitm}, we can get the restricted expressions of the mass of the first resonant state $m_{1}$ with the combination parameter $\gamma$
as
\begin{eqnarray}
m_1&>&(8.8\gamma-14.1)\times{10^{-17}} \text{eV}\label{conditionm1},\\
m_1&\lesssim&(1.5\gamma-2.4)\times{10^{-28+0.2\gamma}}\text{eV}\label{conditionm2}.
\end{eqnarray}
The shadow regions of Fig.~\ref{canshutu2} show the available ranges of the parameters $k$ and $m_{1}$, respectively. From Fig.~\ref{csxianzhik}, we can see that only if $\gamma>57$, the two restricted conditions \eqref{conditionkgamma} and \eqref{conditionM} of $k$ could be satisfied, which means that the parameter $\gamma$ has a lower limit. Correspondingly, Fig.~\ref{csxianzhim1} shows there is a lower limit for the first resonant state mass $m_{1}$, i.e., $m_{1}\gtrsim{10^{-15}}\text{eV}$.

 \begin{figure}[htb]
 \subfigure[$k$]{\label{csxianzhik}
\includegraphics[width=0.225\textwidth]{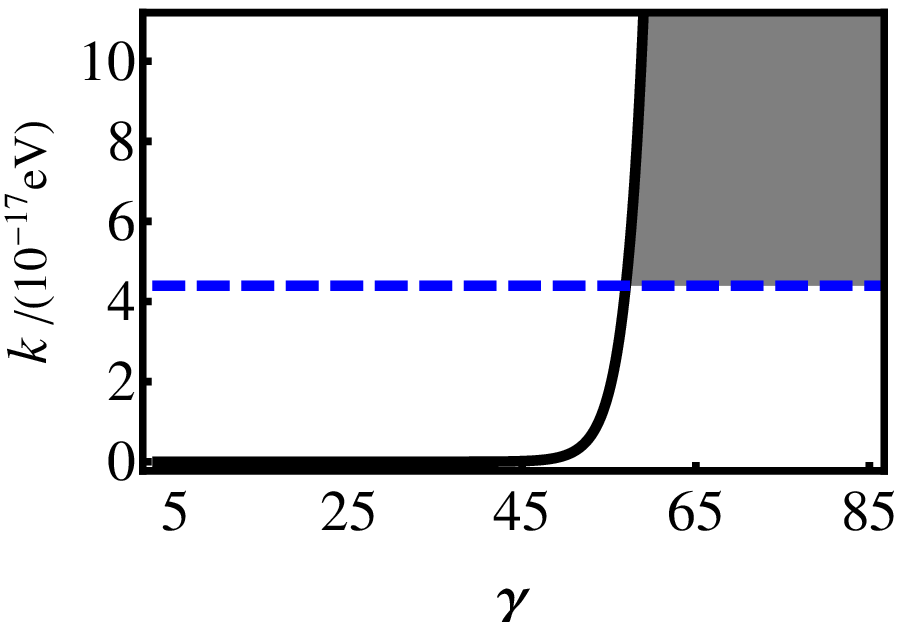}}
\subfigure[$m_{1}$]{\label{csxianzhim1}
\includegraphics[width=0.225\textwidth]{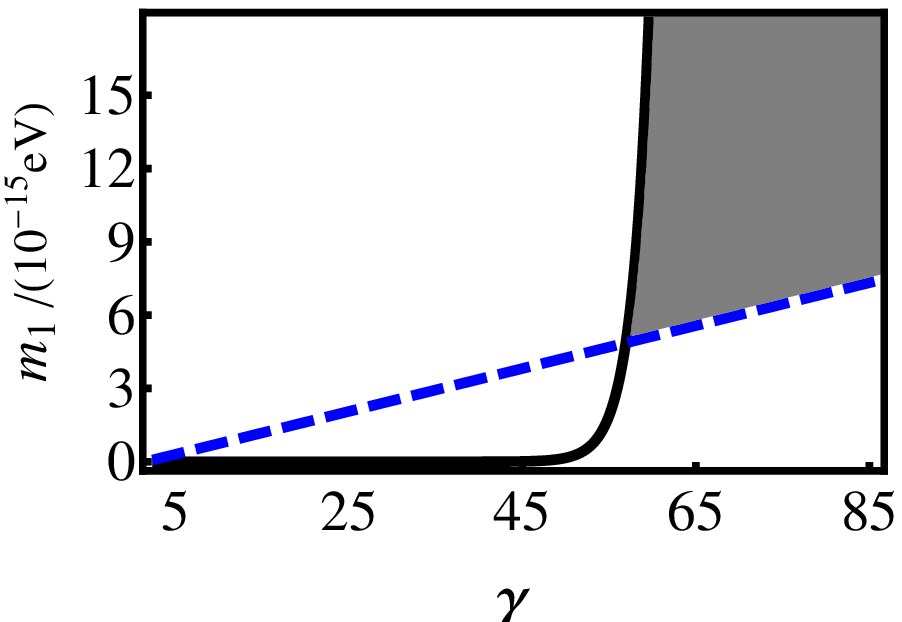}}
\caption{The limit range of the scale parameter $k$ and the mass of the first resonant state $m_{1}$. The blue lines are restrictions from the five-dimensional fundamental scale $M_{*}$ should larger than $13$ TeV and the black lines are restrictions from the  lifetime of the first resonant state should be on the magnitude of the universe life.}
\label{canshutu2}
\end{figure}

\begin{table*}[htb]
\begin{center}
\begin{tabular}{||c|c|c|c|c|c||}
\hline
$\gamma$ & $n$  & $m_{n}$ &  $P$ & $\Gamma$ & $\tau$    \\
\hline \hline
&1& 4.0917 & 0.9474 & $1.0998\times 10^{-7}$ & $9.0924\times 10^{6}$ \\ \cline{2-6}
\raisebox{2.3ex}[0pt]{10}
&2 & 5.1461 & 0.3027 & $2.911\times 10^{-2}$ & $34.3351$ \\ \cline{2-6}
\hline\hline
&1 & 5.1822 & 0.9973 & $1.5778\times 10^{-8}$ & $8.6372\times 10^{7}$    \\ \cline{2-6}
&2 &6.8525 & 0.9032 & $1.4593\times 10^{-6}$& $6.8523\times 10^{5}$  \\ \cline{2-6}
\raisebox{2.3ex}[0pt]{15}
&3& 7.6902 & 0.2839 & $6.5012\times 10^{-3}$& $1.5382\times 10^{2}$ \\ \cline{3-6}
\hline\hline
&1 & 6.8457 & 0.9997 & $1.09557\times 10^{-9}$& $9.1277\times 10^{8}$ \\ \cline{2-6}
&2 & 9.3884 & 0.9468 & $1.598\times 10^{-7}$& $6.2589\times 10^{6}$ \\ \cline{2-6}
&3& 10.9936 & 0.9189 & $9.0939\times 10^{-6}$& $1.0996\times 10^{5} $  \\ \cline{2-6}
\raisebox{2.3ex}[0pt]{25}
&4 & 12.1288 & 0.7774 & $8.2448\times 10^{-4}$&$1.2129\times 10^{3}$ \\ \cline{2-6}
&5& 12.7800 & 0.2456 & $1.9736\times 10^{-3}$& $5.0668\times{10^{2}}$ \\ \cline{2-6}
\hline
\end{tabular}\\
\caption{The influence of combination parameter $\gamma$ on the mass spectrum $m_n$, the relative probability $P$, the width of mass $\Gamma$, and the lifetime $\tau$ of the KK resonances. The scale parameter $ {k}$ is set as $ {k}=1$. }
\label{Table2}
\end{center}
\end{table*}

\begin{table*}[htb]
\begin{tabular}{||c|c|c|c|c|c|c|c|c||}
\hline
$k$ & $n$ &  $m_{n}$ & $m_{n}/k$ & $P$ & $\Gamma$&$\Gamma/k$ & $\tau$ & $\tau*k$    \\
\hline \hline
   &1 & 1.3681 & 6.8405 & 0.9999 & $2.1928\times 10^{-10}$& $1.0964\times 10^{-9}$ & $4.5602\times 10^{9}$&  $9.1204\times 10^{8}$\\ \cline{2-9}
   &2  & 1.8852 & 9.4262 & 0.9987 & $2.6522\times 10^{-8}$ &$1.3261\times 10^{-7}$ & $3.7704\times 10^{7}$&$7.5408\times 10^{7}$  \\ \cline{2-9}
   &3 & 2.1962 & 10.9813 & 0.9872 & $1.8213\times 10^{-6}$ & $9.1065\times 10^{-6}$ &$5.4904\times 10^{5}$&$1.0981\times 10^{5}$  \\ \cline{2-9}
\raisebox{2.3ex}[0pt]{0.2}
  &4 & 2.4306 & 12.1529 & 0.8212 & $1.6546\times 10^{-4}$ &$8.2731\times 10^{-4}$ & $6.0767\times 10^{3}$ &$1.2153\times 10^{3}$ \\ \cline{2-9}
  &5  & 2.5470 & 12.7350 & 0.3192 & $3.5343\times 10^{-3}$ & $1.7672\times 10^{-3}$ &$2.8294\times 10^3$&$5.6588\times 10^{2}$ \\ \cline{2-9}
\hline\hline
&1& 13.6982 & 6.8493 & 0.9993 & $2.1902\times 10^{-9}$& $1.0951\times 10^{-9}$ & $4.5657\times 10^{8}$& $9.1314\times 10^{8}$    \\ \cline{2-9}
&2  & 18.7361 & 9.3680 & 0.9923 & $2.6686\times 10^{-7}$& $1.3343\times 10^{-7}$& $3.7472\times 10^{6}$& $7.4944\times 10^{6}$  \\ \cline{2-9}
\raisebox{2.3ex}[0pt]{2}
&3 & 22.0077 & 11.0039 & 0.9361 & $1.8175\times 10^{-5}$& $9.0875\times 10^{-6}$& $5.5021\times 10^{4}$& $1.1004\times 10^{5}$ \\ \cline{3-9}
&4  & 24.2305 & 12.1141 & 0.8164 & $1.6507\times 10^{-3}$& $8.2535\times 10^{-4}$& $6.0578\times 10^{2} $& $1.2116\times 10^{3}$\\ \cline{2-9}
&5  & 25.6334 & 12.8167 & 0.3082 & $3.5109\times 10^{-3}$ & $1.7554\times 10^{-3}$& $2.8482\times 10^{2}$& $5.6964\times 10^{2}$\\ \cline{2-9}
\hline\hline
&1  & 34.1950 & 6.8391 & 0.9991 & $5.4417\times 10^{-9}$& $1.0883\times 10^{-9}$& $1.8377\times 10^{8}$& $9.1883\times 10^{8}$ \\ \cline{2-9}
&2  & 47.2038 & 9.4404 & 0.9986 & $7.4164\times 10^{-7}$& $1.4833\times 10^{-7}$& $1.3484\times 10^{6}$& $6.7419\times 10^{6}$ \\ \cline{2-9}
&3   & 54.8917 & 10.9784 & 0.9822 & $4.3735\times 10^{-5}$& $8.7471\times 10^{-6} $ & $2.8654\times 10^{4}$& $1.4327\times 10^{5}$ \\ \cline{2-9}
\raisebox{2.3ex}[0pt]{5}
&4   & 60.8046 & 12.1611 & 0.8203 & $4.1115\times 10^{-3}$& $8.2234\times 10^{-4}$&$2.43\times 10^{2}$& $1.2166\times 10^{3}$ \\ \cline{2-9}
&5  & 63.6192 & 12.7238 & 0.3290  & $7.8591\times 10^{-3}$& $1.5718\times 10^{-3}$&$1.2724\times 10^{2}$& $6.3227\times 10^{2}$ \\ \cline{2-9}
\hline
\end{tabular}\\
\caption{The influence of scale parameter $k$ on the mass spectrum $m_n$, the relative value $m_n/k$, the relative probability $P$, the width of mass $\Gamma$, the the relative width as $\Gamma/k$, the lifetime $\tau$, and the relative $\tau*k$ of the KK resonances. The combination parameter $\gamma$ is set as $\gamma=25$.}
\label{Table1}
\end{table*}


\section{Conclusion}
We generalized the geometrical coupling mechanism in order to localize a five-dimensional vector field on RS-like thick branes. The key feature of the mechanism is to introduce two mass terms of the vector field, which are proportional to the five-dimensional Ricci scalar and the Ricci tensor, respectively. We decomposed the vector field $A_{M}$ into three parts: the vector part $\hat{A}_{\mu}$, the scalar part $\phi$ and $A_{5}$. With the transverse condition $\partial_\mu\hat{A}^\mu=0$, we got a decoupled action of $\hat{A}_{\mu}$. We find that when the two parameters $\alpha$ and $\beta$ in the action (\ref{vectorAction}) satisfy the relation $\beta=-1-8\alpha\pm\sqrt{1+12\alpha}$, the effective potential $V_v(z)$ of the vector KK modes can be expressed as $V_v(z)=\gamma^2(\partial_{z}A)^2
+\gamma\partial^2_{z}A$ with $\gamma=\frac{3}{2}\pm\sqrt{1+12\alpha}$ and the tachyonic KK modes of $\hat{A}_{\mu}$ can be excluded. For $\gamma>1/2$, the zero mode of $\hat{A}_\mu$ can be localized on the brane.

Then, we investigated the resonances of the vector field by using the relative probability method and considered the possibility of these resonances as one of the candidates of dark matter. We analyzed the influence of the parameters $k$ and $\gamma$ on the resonant behavior. We found that, only for $\gamma>3$, the massive resonant KK modes could exist. Both the two parameters affect the height of the potential and hence the vector resonances. The number of the resonant states only increases with the parameter $\gamma$. We also considered scaled lifetime $\bar{\tau}$ and the scaled mass $\bar{m}_{1}$ of the first resonant state. We found that both the scaled mass $\bar{m}_{1}$ and the scaled lifetime $\text{log}(\bar{\tau})$ can be fitted by a linear function of $\gamma$ approximately. In order to view the first resonant vector KK state as dark matter, its lifetime should be long enough as the age of our universe. This would introduce some constrains on the parameters $k$ and $\gamma$ as well as the mass of the first resonance, i.e., $k\gtrsim{10^{-17}} \text{eV}$, $\gamma >57$, and $m_{1}\gtrsim {10^{-15}}\text{eV}$.

Note: When we finish this work, we find another work \cite{2001.01267} that also considered the same localization mechanism (\ref{vectorAction}) for the vector filed.

\section*{Acknowledgments}
This work was supported by the National Natural Science Foundation of China (Grants No.~11875151, No.~11705070, and No.~11522541), and the Fundamental Research Funds for the Central Universities (Grants No.~lzujbky-2018-k11, and No.~lzujbky-2019-it21).

\end{document}